\documentclass[%
 reprint,
 amsmath,amssymb,
 aps,
floatfix,
]{revtex4-2}

\usepackage{graphicx}
\usepackage{dcolumn}
\usepackage{bm}
\usepackage{hyperref}
\usepackage{braket}
\usepackage[utf8]{inputenc}
\usepackage[english]{babel}
\usepackage{amsmath,calc} 
\usepackage{mathtools} 
\usepackage{todonotes}

\graphicspath{{./Figures/}}

\begin{document}


\title{Unified description of saturation and bistability of intersubband transitions \\ in the weak and strong light-matter coupling regimes}

\author{Mathieu Jeannin}

\author{Jean-Michel Manceau}

\author{Raffaele Colombelli}
\email{E-mail: raffaele.colombelli@u-psud.fr}

\affiliation{Centre de Nanosciences et de Nanotechnologies (C2N), CNRS UMR 9001, Universit\'e Paris-Saclay, 91120 Palaiseau, France}

\date{\today}

\begin{abstract}
We propose a unified description of intersubband absorption saturation for quantum wells inserted in a resonator, both in the weak and strong light-matter coupling regimes. We demonstrate how absorption saturation can be engineered. In particular we show that the saturation intensity increases linearly with the doping in the strong coupling regime, while it remains doping independent in weak coupling. Hence, countering intuition, the most suitable region to exploit low saturation intensities is not the ultra-strong coupling regime, but is instead at the onset of the strong light-matter coupling. We further derive explicit conditions for the emergence of bistability. This work sets the path towards yet unexisting ultrafast mid-infrared semiconductor saturable absorption mirrors (SESAMs) and bistable systems. As an example, we show how to design a mid-infrared SESAM with a three orders of magnitude reduction in saturation intensity, down to $\approx 5$ kWcm$^{-2}$.
\end{abstract}

\maketitle

Saturation of the light-matter interaction is a general feature of material systems, be they atoms or semiconductors.~\cite{Boyd2008} 
In semiconductors, the possibility of judiciously controlling saturation phenomena is of great importance for fundamental physics as well as applications.

A seminal example is the development of the semiconductor saturable absorption mirror (SESAM)~\cite{Keller:92} based on interband transitions in semiconductor quantum wells (QW) that revolutionized the field of ultra-fast lasers in the visible/near-IR spectral range. Ultra-fast lasers based on SESAMs now find applications in several domains, and even in quantum phenomena~\cite{Willenberg2019}.
Saturable absorption can lead to optical bistability~\cite{Boyd2008} that manifests itself as an hysteresis cycle in the input-\textit{vs}-output characteristic of the system. Besides fundamental interests, it has been proposed as a resource for optical information processing. Proposals~\cite{TrediBista} and demonstrations~\cite{giacobino2004} exist in the domain of exciton polaritons.

A saturable absorber exhibits an absorption coefficient $\alpha$ that depends on the incident intensity $I$ as:
\begin{equation}
\alpha = \frac{\alpha_0}{1+I/I_{sat}}    
\label{eq:saturable}    
\end{equation}
$\alpha_{0}$ is the absorption at low incident power, and $I_{sat}$ is the saturation intensity. At $I_{sat}$, the absorption is 50\% of the low-intensity value. In general, $I_{sat}$ depends on the characteristics of the system: lifetimes, oscillator strengths, operating frequency.
The only way to engineer it is to act on the system lifetimes. Fortunately, the saturation intensities of systems based on \textit{interband} transitions in semiconductors are naturally favorable for applications in the visible/near-IR spectral ranges.

The case is different for \textit{intersubband transitions} (ISBT) in QWs, the backbone of semiconductor-based mid-IR optoelectronics (quantum cascade lasers (QCLs), QW infrared detectors) ~\cite{helm1999basic,BaranovBook}.
Considering typical experimental configurations, the classical theory yields a saturation intensity in the absence of a cavity/resonator~\cite{rosencher2002optoelectronics}:
\begin{equation}
I_{sat}^{0}=
\frac{\hbar^2 \varepsilon_0 c n_{opt}}{2 e^2 \tau_{12} T_2 |\bra{1}z\ket{2}|^2 } =
\frac{\varepsilon_0 c n_{opt}}{2\tau_{12} T_2 |(D_{12}/\hbar)|^2 }
\label{eq:sat-rosencher}
\end{equation}
where $n_{opt}\approx3.3$ is an average index of refraction of the active region, $\tau_{12}\approx0.8$~ps is upper state lifetime, $T_{2}\approx60$~fs is the dephasing time, $|\bra{1}ez\ket{2}| = D_{12}\approx e\times 2$~nm is the electric dipole matrix element. 
In the mid-IR ($\lambda=10\ \mu$m), $I_{sat}^{0}\approx1$~MW/cm$^{2}$, as confirmed by a vast body of literature~\cite{Seilmeier_IntersubbandBleaching,Julien-Saturation,Vodopyanov_1997}. 
This very high value, that notably does not depend on the doping, explains why saturable absorbers, SESAM mirrors, bistable systems are missing from the current toolbox of mid-IR opto-electronic devices: they could only be used with extremely high power laser sources and are incompatible with the typical output power levels of QCLs. 
%

The nature of the absorption saturation for an ISB system is radically modified when operated in the strong light-matter coupling regime. In this regime, new quasi-particles emerge (intersubband polaritons) with peculiar properties that motivate the recent flurry of activity. From their initial demonstration \cite{Dini2003}, a non-exhaustive list of current research includes electrically and optically pumped devices towards demonstrating bosonic final state stimulation \cite{DeLiberato_StimulatedScattering, Delteil_OpticalPhononScattering, Zanotto-Saturation, Colombelli2015, Manceau2018}. The ultra-strong coupling regime where a sizeable number of virtual photons is populating the ground state of the system has been widely investigated \cite{todorov2010ultrastrong, Scalari2012, Delteil2012}, along with the possibilities to reach this regime with fewer electrons \cite{Malerba2017, Keller2017, Jeannin_Ultrastrong} and its non-adiabatic modulation to emit non-classical states of light \cite{ciuti2005quantum, Gunter2009}. More in-line with the present proposal, devices such as amplitude modulators are under development \cite{Lee2014, pirotta2020ultrafast} and novel approaches to study transport in polaritonic systems are emerging \cite{Vigneron2019, Limbacher_RTD_2020}.

In this letter, we propose a new strategy to exploit absorption saturation of ISB transitions and trigger their non-linear behaviour with moderate pumping powers. The crucial aspect of this work is that we rely on the collapse of the light-matter coupling resulting from the saturation of the transition to produce a strong spectral feature. We show that the key to reduce the saturation intensity is to engineer the crossover between the weak (WCR) and the strong (SCR) coupling regimes so that it appears at a low doping. In contrast with the absorption saturation of interband excitations, both in the WCR and the SCR, we do not rely on Pauli blocking like traditional SESAM mirrors \cite{Keller:92}, nor on polariton-polariton interactions that dominate the non-linear response of exciton polaritons \cite{TrediBista, giacobino2004}. We develop a unified analytical formalism for absorption saturation covering both the WCR and the SCR. We provide a set of simple analytical formulas that permit to assess the saturation power levels for a given ISB system. 
Countering intuition, we show that in the strong light-matter coupling regime, $I_{sat}$ \emph{increases} linearly with the light-matter coupling strength (i.e. with the doping), while in the WCR it remains doping-independent.  To illustrate the importance of this finding, we present the design of a SESAM with extremely low saturation intensity and further derive the conditions to reach bistability. This formalism could apply to other systems hosting collective excitations in the strong coupling regime like vibrational absorbers with an adequate description of the population saturation of the excited state. 
\par

Let us start by an intuitive description of the saturation intensity in both coupling regimes. 
We model the saturation with a simple rate equation approach. The steady-state expression of the surface charge density $n_2$ in the excited subband of a system of doped QWs with surface charge density per QW $n_s$ is:
\begin{equation}
n_2 = \frac{I}{\hbar\omega}\tau_{12}\mathcal{A}^{qw}(\Delta n,\omega) =  \frac{I}{N_{qw}\ \hbar\omega}\tau_{12}\mathcal{A}^{ISB}(\Delta n,\omega)
\label{eq:n2population_severalQWs}
\end{equation}
$I$ is the incident intensity per unit surface, $\tau_{12}$ is the upper state lifetime, $N_{qw}$ is the number of QWs, $\mathcal{A}^{qw}(\Delta n,\omega)$ is the absorption per QW, $\mathcal{A}^{ISB}(\Delta n,\omega)$ is the \textit{total} absorption of the ISB system, and $\Delta n = n_1 -n_2$ is the population difference. 
The saturation condition is defined as $n_2 = \frac{n_s}{4} \Longleftrightarrow \Delta n = \frac{n_s}{2} $
The saturation intensity is thus derived from: 
\begin{equation}
\frac{n_s}{4} = \frac{I_{sat}}{N_{qw}\ \hbar\omega}\tau_{12}\mathcal{A}^{ISB}(\frac{n_s}{2},\omega)
\label{eq:saturation}
\end{equation}

Remarkably, as long as the absorption from the ISB system can be written as a \textit{linear} term in the doping density $n_s$, $I_{sat}$ is doping independent. While this is explicit in the bulk saturation intensity $I_{sat}^{0}$ from eq.~\eqref{eq:sat-rosencher}, this is in fact a general feature that also applies in the presence of a cavity when the system operates in the WCR. The absolute value $I_{sat}$ can be dramatically improved using a cavity, but it remains independent of the doping density because in the WCR, the absorption from the ISB system is still a \textit{linear} term in $n_s$. 
Instead in the SCR, the absorption from the ISB system is independent of the doping because any further increase in the charge concentration results in an increase of the Rabi splitting (see Supplementary Material). It is then straightforward to infer from Eq.~\eqref{eq:saturation} a saturation intensity that increases linearly with $n_s$ in SCR, in stark contrast with the WCR case.

\begin{figure}[!hbt]
\centering
\includegraphics[width=\linewidth]{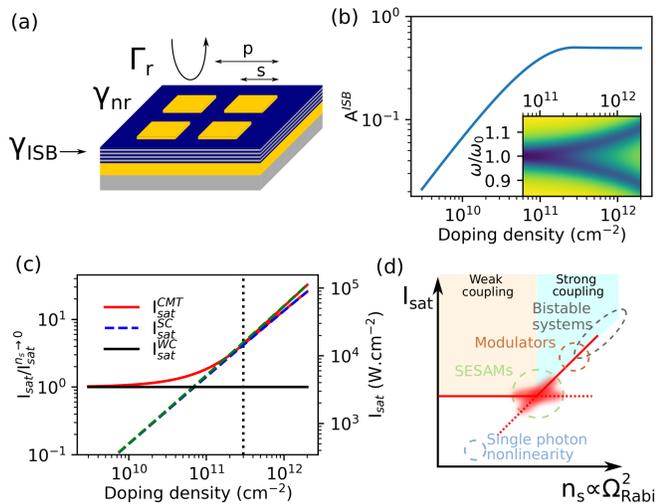}
\caption{(a) Sketch of the QWs embedded in a microcavity array with relevant parameters. (b) Intersubband absorption $\mathcal{A}^{ISB}(n_s,\omega)$ as a function of $n_s$. 
Inset: Reflectivity of the system as a function of $n_s$. 
(c) Saturation intensity normalized to $I_{sat}^{WC}$ (left axis) or following our proposal (right axis) as a function of $n_s$. 
The black solid line is the asymptotic saturation intensity in the WCR $I_{sat}^{WC}$, and the two dashed lines are the 
asymptotic saturation intensity $I_{sat}^{SC}$ in the SCR corresponding to the lower (blue) and upper (green) polaritons.
The red solid line is the analytical solution $I_{sat}^{\pm,CMT}$. The black dotted line shows the proposed design for low saturation SESAMs. (d) Sketch of the two operating regimes and potential devices as a function of $n_s$.
}
\label{fig:fig2}
\end{figure}

To gain further insight into the physic at play, we use the temporal coupled mode theory (TCMT) formalism to derive a simple set of analytical expressions and unify both coupling regimes. 
In this framework, the cavity mode and the ISB system are modelled as oscillators with characteristic parameters 
$(\omega_i, \gamma_i, \Gamma_i)$ representing respectively their natural oscillation frequency, non-radiative, and radiative dampings as schematically depicted on Fig.~\ref{fig:fig2}(a).
In the case of the ISB system, the radiative coupling of the ISB transition to free-space radiation is negligible \cite{alpeggiani2014semiclassical, Jeannin-abseng}.
The coupled system's response to an external excitation is given by:
\begin{align}
    \frac{\mathrm{d}a_{isb}}{\mathrm{d}t} =& (i\omega_{isb}-\gamma_{isb})a_{isb}
        + i\Omega_{Rabi} a_{c}  \label{eq:TCMTsystem_eq1} \\
    \frac{\mathrm{d}a_{c}}{\mathrm{d}t} = & (i\omega_{c}-\gamma_{nr}-\Gamma_{r})a_{c} 
         + i\Omega_{Rabi} a_{isb} + \sqrt{2\Gamma_{r}}s^+ \\
    s^- = & -s^+ + \sqrt{2\Gamma_{r}} a_{c} \\
    n_2 = & 2\tau_{12}\gamma_{isb}\frac{\left|a_{isb}\right|^2}{N_{qw}\hbar \omega}   \label{eq:population_CMT}
\end{align} 
with $a_i$ the amplitude of oscillator $i$ referencing either the ISB system ($i=isb$) or the cavity ($i=c$), 
$\gamma_{isb}$ and $\gamma_{nr}$ the non-radiative decay rate respectively of the ISB transition and of the cavity, 
$\Gamma_{r}$ the cavity radiative decay rate, $s^+$ and $s^-$ the amplitudes of the incoming and reflected fields.
The coupling rate between the two oscillators is the vacuum Rabi frequency, which depends on the population difference:
\begin{equation}
    \Omega_{Rabi}^2 = f_w \frac{\Delta n ~e^2}{4 \varepsilon \varepsilon_0 m^{\ast} L_{qw}} = a ~\Delta n \label{eq:Rabi}
\end{equation}
where $f_w$ is the filling fraction of the QWs material in the active region. 
Equations \eqref{eq:population_CMT} and \eqref{eq:Rabi} ensure the self-consistency of the set of CMT equations.
We assume an ISB linewidth independent of the doping density. While this is true for most of the values used throughout this manuscript, at very large doping densities the ISB linewidth can increase \cite{Warburton_1998}. Careful positioning of the dopants and growth interruption techniques can compensate for this increase \cite{growth-interruptions}. We thus keep $\gamma_{isb}$ constant.
Our formalism applies to the case when only the lowest subband is occupied and only the fundamental transition plays a role.
Solving this system in the harmonic regime ($s^+=\mathrm{e}^{i\omega t}$) we derive the reflectivity of the coupled system $R=\left|\frac{s^-}{s^+}\right|^2$ (see Suppl. Mat.) and the absorption of the sole ISB system:

\begin{align} 
    \mathcal{A}^{ISB} 
        =& \frac{4\gamma_{isb}\Gamma_{r} \Omega_{Rabi}^2 }
            {\splitfrac{\left[\gamma_{isb}(\gamma_{nr} +\Gamma_{r})-((\omega-\omega_0)^2-\Omega_{Rabi}^2)\right]^2}{+(\gamma_{isb}+\gamma_{nr}+\Gamma_{r})^2(\omega-\omega_0)^2}}
            \label{eq:AISB}  
\end{align}

assuming without loss of generality that the cavity and ISB system are resonant ($\omega_{isb}=\omega_{c}=\omega_0$). 

Using these equations, we study the response of the cavity-coupled ISB system as a function of the doping density $n_s$ for a weak probe beam. 
The maximum ISB absorption following eq.~\eqref{eq:AISB} is shown in Fig.~\ref{fig:fig2}(b).
Two regimes appear clearly: for low carrier density the absorption increases linearly with $n_s$, 
while for large doping the maximum ISB absorption flattens and becomes doping independent. \\
The transition from the WCR to the SCR produces a clear spectral signature with the energy splitting of the two resonances (Fig.~\ref{fig:fig2}(b), inset).
The expression of the ISB absorption \eqref{eq:AISB} can be simplified in the two limiting cases of very low doping (WCR) or large doping (SCR).
Plugging the asymptotic values of $\mathcal{A}^{ISB}$ into eq.~\eqref{eq:saturation} 
leads to analytical expressions for the saturation intensity:
\begin{align}
    I_{sat}^{WC} =& \frac{\hbar\omega_0\varepsilon\varepsilon_0m^{\ast}L_{qw}N_{qw}}{2\tau_{12}f_w e^2}
        \frac{\gamma_{isb}(\gamma_{nr}+\Gamma_{r})^2}{\Gamma_{r}}
        \label{eq:sat-enhancement-TCMT}\\
    I_{sat}^{SC} =& \frac{n_s\hbar \omega_{\pm}
            N_{qw}}{16\tau_{12}}\frac{(\gamma_{isb}+\gamma_{nr}+\Gamma_{r})^2}{\gamma_{isb}\Gamma_{r}} \label{eq:sat-SC-TCMT}
\end{align}
where $\omega_{\pm}$ represents the polariton frequencies.\\
$I_{sat}^{WC}$ (black solid lin in Fig. \ref{fig:fig2}(c)) represents the limit of low doping: the system operates in the weak coupling regime, and is evidently independent of the doping density $n_s$.
Conversely, $I_{sat}^{SC}$ (dashed lines in Fig.~\ref{fig:fig2}c)) corresponds to a system operating in the strong coupling regime: the saturation intensity increases linearly with $n_s$ and $N_{qw}$, confirming our previous intuitive discussion.
\par

We re-arrange the weak coupling limit expression \eqref{eq:sat-enhancement-TCMT} to highlight the key parameters governing the reduction in $I_{sat}$ once the QWs are placed in cavity.
In line with the rest of this work, we assume a MIM (metal-insulator-metal) geometry, so that the electric field is almost constant across the active region:
\begin{align}
    I_{sat}^{WC} & =  f_{osc}\frac{\pi L_{AR}}{(\lambda/n_{opt})} \frac{Q_{r}}{2\cdot Q_{cav}^{{tot}^2}} \cdot I_{sat}^{0} 
         \label{eq:sat-enhancement}
    \end{align}
where $f_{osc}$ is the ISB oscillator strength~\cite{helm1999basic}, $L_{AR}$ is the total thickness of the active region, 
and $Q_{r}$ and $Q_{cav}^{tot}$ are the cavity radiative and total Q-factor, respectively. 
Eq.~\eqref{eq:sat-enhancement} explicitly links the lowest saturation intensity one can achieve in a cavity-coupled system ($I_{sat}^{WC}$) to the cavity-free one. Leaving aside the (fixed) oscillator strength, eq.~\eqref{eq:sat-enhancement} shows that two physical effects lead to a reduction in $I_{sat}$. The first one stems from the compression of the electric field in ultra-subwavelength volumes. The second one is the branching ratio of the cavity quality factors. Its maximization requires both avoiding material losses (large $Q_{nr}$) while retaining the critical coupling condition $Q_{r}=Q_{nr}$ to ensure maximal energy feeding into the system. 

The exact analytical solution is obtained by substituting eq.~\eqref{eq:AISB} in eq.~\eqref{eq:saturation} (red solid line in Fig.~\ref{fig:fig2}(c)), demonstrating the system's behaviour around the crossing point and the broad span of ISB absorption saturation engineering that can be obtained with different doping densities. We summarize these different perspectives in Fig.~\ref{fig:fig2}(d). Ultra-fast modulators relying on SCR~\cite{pirotta2020ultrafast}, as well as harmonic generation with metasurfaces \cite{Lee2014,second-harmonic-surface-1} suit the ultra-strong coupling regime, as they must not easily saturate. The regime of ultra-low power nonlinearities in the mid-IR is an interesting perspective. While this requires very low loss resonators and is beyond the scope of this paper, we speculate that mid-IR on-chip III-V photonics is a suitable platform~\cite{YannisChip}. More interestingly, we anticipate the possibility to design and demonstrate SESAM mirrors. Such device would best fit the onset of SCR, as low $I_{sat}$ can be achieved, retaining at the same time a high reflectivity in the saturated state.

We rely on this formalism to devise a semiconductor-based saturable system operating with low $I_{sat}$ at mid-IR wavelengths ($\omega_0=30$~THz, $\lambda=10\ \mu$m). A superficial inspection of Fig.~\ref{fig:fig2}(c) suggests that the lower the doping, the better. However in the case where the system operates in the weak coupling regime, the main practical limitation arises from the overlap factor $f_w$ (see eq.~\eqref{eq:sat-enhancement-TCMT}) that must remain elevated. In the mid-IR, overlap factors around unity are obtained only using MIM cavities, while dielectric cavities result in very low $f_w$ values. In turn, saturation would be almost undetectable in a MIM cavity operating in the WCR, since the losses are dominated by the cavity, and the saturated/non-saturated reflectivity spectra would be almost identical. It is then necessary to operate in the SCR, where saturation leads to clear spectral changes. 
Equation~\eqref{eq:sat-SC-TCMT} clarifies that the system must operate at the onset of SC, embed the lowest possible number of QWs, and the lowest possible sheet doping per QW allowing operation in the SCR. 
We consider a system of 4 periods of GaAs/Al$_{0.33}$Ga$_{0.67}$As (8.3 nm/14 nm), with barriers $\delta$-doped to $n_s=3~10^{11}$ cm$^{-2}$, embedded in metal-metal patch-cavity resonators (Fig.~\ref{fig:fig2}(a))~\cite{balanis-book,todorov2010ultrastrong,second-harmonic-surface-1,hakl2020ultrafast,Cortese_2020}. The structure thickness is $\approx 140$~nm, and the overlap factor is $f_w \approx 0.24$.

\begin{figure}[!hbt]
\centering
\includegraphics[width=\linewidth]{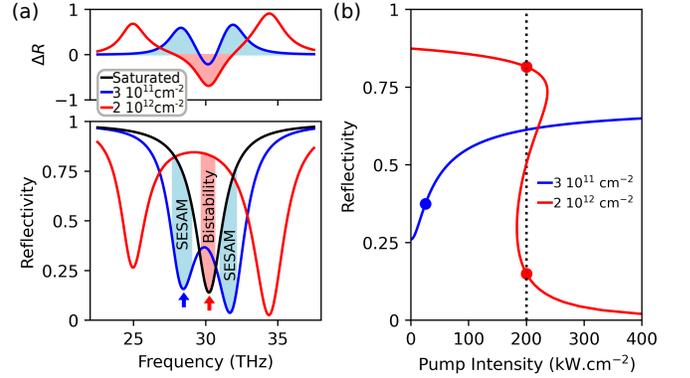}
\caption{
(a) Bottom: FEM reflectivity simulations for two different doping densities. 
Blue, red: low incident power, the polaritonic states are present.
Black: high incident power, only the cavity resonance is visible.
The arrows indicate the pump frequency used in panel (b).  
Top: Differential reflectivity $\Delta R$.
(b) Reflectivity of the two design samples, SESAM (blue, $n_s=3~10^{11}$~cm$^{-2}$) and bistable (red, $n_s=2~10^{12}$~cm$^{-2}$) as a function of pump intensity obtained from TCMT at the frequencies shown with arrows in (a). 
}
\label{fig:fig3}
\end{figure}

We numerically calculate the system reflectivity in strong coupling with a patch cavity mode at low incident power
using finite element method (FEM) simuations (Fig.~\ref{fig:fig3}(a) blue curve, lower panel): 
the UP and LP modes are clearly visible. The black curve shows instead the saturated system ($I> I_{sat}$). The top panel shows the differential reflectivity. 
The regions highlighted in light blue correspond to an absorption that \textit{decreases} with incident power: they suit the development of SESAM mirrors.
The region highlighted in red correspond to an absorption that \textit{increases} with incident power: we will discuss later how under specific conditions, a bistable system could be implemented here.
\par
From the FEM reflectivity spectra, we can extract the relevant system parameters to use as inputs into the TCMT, respectively $\gamma_{nr} = 0.02\omega_c$ and $\Gamma_{r} = 0.02\omega_c$. The non-radiative damping rate of the ISB system is set with typical intersubband absorption $Q$ factors, $\gamma_{isb} = 0.05\omega_{isb}$. We calculate $I_{sat}$ as a function of the doping for these sample parameters, as reported in Fig.~\ref{fig:fig2}(c). 
We obtain a WCR limit value of $I_{sat}^{WC} = 3.45$~kW/cm$^2$. Considering the previously chosen doping level of $3~10^{11}$ cm$^{-2}$ per QW, we get $I_{sat}^{SC}=14.7$~kW/cm$^2$, with a Rabi frequency $\Omega_{Rabi} \approx 3$~THz. Note that the saturation intensity of the same active region in absence of a cavity is around $I_{sat}^{0}=1.1~$MW.cm$^{-2}$. Our approach allows at least an eighty-fold reduction of the saturation intensity.

This result, which already represents a record-low value for mid-IR absorption saturation in an ISB system, is an upper limit. Arranging microcavities in an array enhances their absorption cross-section as they can collect photons over an area greater than their physical area \cite{second-harmonic-surface-1,Seok-photonharvest,Todorov-Thz-enhanc}. Numerically optimizing the array absorption ($s=1.2\ \mu$m; $p=2.2\ \mu$m), the active surface covers 30\% of the unit cell. Hence, we estimate a corrected saturation intensity $I_{sat}^{enh} = 0.3\cdot I_{sat}^{SC}\approx 4.4$ kW/cm$^2$.
 %
%

So far, we have only focused on quantifying and engineering $I_{sat}$. Equations \eqref{eq:TCMTsystem_eq1}-\eqref{eq:Rabi} also allow to derive the steady-state response of the system to a variable pump power. 
This is particularly useful to predict experimental data on the observation of non-linear optical properties of the coupled system. 
We follow the method developed for exciton polaritons in Ref.~\cite{giacobino2004}, and express the reflected beam intensity, from the population of the excited state using the TCMT equations. It is best demonstrated by writing the intensity-dependent excited state population as a function of the coupled system's parameters. For simplicity we assume without loss of generality that the pump frequency is set to $\omega = \omega_0$:
\begin{equation}
    I(n_2) =  \frac{n_2\hbar\omega_0 N_{qw}\left[ 2a n_2 -(\gamma_{isb}(\gamma_{nr}+\Gamma_{r}) - a n_s) \right]^2}{4 \tau_{12} \gamma_{isb} \Gamma_{r} a (n_s-2n_2)}
    \label{eq:bistability}
\end{equation}
The derivation and its extension to the detuned pump ($\omega \neq \omega_0$) used to describe the SESAM operation are presented in Supplementary Material.
Under certain conditions, equation \eqref{eq:bistability} admits several solutions: multiple values of $n_2$ can satisfy the equation for a given pump intensity $I$, the system is then said to exhibit \emph{optical bistability}. The emergence of bistability in ISB systems has been previously theoretically approached in Refs.~\cite{Zanotto-Saturation,helm-bistab,Zaluzny-sat} either through charge transfer or absorption saturation.

The observation of bistability in the system requires very large changes in absorption, and is thus usually predicted at prohibitively large pumping intensities \cite{helm-bistab}. We show here that it can emerge at lower pumping powers. From eq.~\eqref{eq:bistability}, the minimal doping necessary to observe bistability as a function of the system's damping rates is:
\begin{equation}
    n_s > \frac{32 \varepsilon \varepsilon_0 m^{\ast} L_{qw} \gamma_{isb}(\gamma_{nr}+\Gamma_{r})}{f_w e^2}
    \label{eq:bistab-condition}
\end{equation}
Using the previous parameters, this critical value is around $n_s \geq 1.1~10^{12}$~cm$^{-2}$.  
To illustrate this regime of operation, we explore a sample with a doping level $n_s=2~10^{12}$~cm$^{-2}$. FEM calculations of the bistable sample are presented in Fig.~\ref{fig:fig3}(a) (red lines). 
Figure~\ref{fig:fig3}(d) shows the reflected beam intensity when pumping the SESAM device at $\omega_{-}\approx28$~THz, and the bistable device at $\omega_0=30$~THz. We evidence a bistable behaviour for pump intensities around 200~kW/cm$^2$ (red curve) with a classical s-shape reflectivity curve. The saturable absorber sample with a doping level $n_s=3~10^{11}$~cm$^{-2}$ pumped at the lower polariton frequency (blue) shows a classical saturation behaviour, with $I_{sat}$ marked as the blue circle.

Reminding the definition of plasma frequency~\cite{zaluzny1999coupling}, Eq.~\eqref{eq:bistab-condition} can be re-cast in a universal form, linking it to modern cavity (quantum) electrodynamics concepts:
\begin{equation}
    \frac{\Omega_{Rabi}}{\omega_{isb}} \cdot \frac{\Omega_{Rabi}}{\omega_{cav}}
    >
    \frac{2}{Q_{isb}\ Q_{cav}^{tot}} \iff C > 8
    \label{eq:short-bistab-condition}
\end{equation}
Where $C=\Omega_{Rabi}^2/\gamma_{isb}(\gamma_{nr}+\Gamma_{r})$ is the cooperativity. Remarkably, this result is extremely similar to the bistability criterion reported for atomic ensembles in interferometers and cavities \cite{Drummond_bista, Grant:82}. Equation~\eqref{eq:short-bistab-condition} explicitly spells out the physical parameters governing bistability in an ISB polariton system, which has never been experimentally reported.

In conclusion, we have demonstrated that the nature of absorption saturation for an ISB system changes radically when the system transitions from the weak to the strong light-matter coupling regime. In particular, in the SCR the saturation intensity becomes doping dependent. This brings a new degree of freedom, and reveals that the true nature of absorption saturation emerges when we encompass the full range of possible light-matter couplings, from the weak up to the strong coupling regime. Our work shows the richness of possibilities to tailor ISB absorption in strong coupling with respect to the weak coupling case. We have further provided design rules to enable record-low threshold intensity saturable and bistable mid-IR systems.

During the revision of this manuscript, we came aware of new work showing ISB polariton saturation in metasurfaces \cite{Mann_Metasurface_2021}. Our theory quantitatively reproduces those results.
\begin{acknowledgments}
We  thank I. Carusotto for thorough reading of the manuscript, and F.H. Julien, M. Helm, L. Andreani, A. Bousseksou, S. Pirotta,  for discussions.
We acknowledge financial support from the European Union FET-Open Grant MIRBOSE (737017) and from the French National Research Agency (project TERASEL, ANR 18-CE24-0013 and project SOLID, ANR 19-CE24-0003). 
\end{acknowledgments}


\begin{thebibliography}{46}%
\makeatletter
\providecommand \@ifxundefined [1]{%
 \@ifx{#1\undefined}
}%
\providecommand \@ifnum [1]{%
 \ifnum #1\expandafter \@firstoftwo
 \else \expandafter \@secondoftwo
 \fi
}%
\providecommand \@ifx [1]{%
 \ifx #1\expandafter \@firstoftwo
 \else \expandafter \@secondoftwo
 \fi
}%
\providecommand \natexlab [1]{#1}%
\providecommand \enquote  [1]{``#1''}%
\providecommand \bibnamefont  [1]{#1}%
\providecommand \bibfnamefont [1]{#1}%
\providecommand \citenamefont [1]{#1}%
\providecommand \href@noop [0]{\@secondoftwo}%
\providecommand \href [0]{\begingroup \@sanitize@url \@href}%
\providecommand \@href[1]{\@@startlink{#1}\@@href}%
\providecommand \@@href[1]{\endgroup#1\@@endlink}%
\providecommand \@sanitize@url [0]{\catcode `\\12\catcode `\$12\catcode
  `\&12\catcode `\#12\catcode `\^12\catcode `\_12\catcode `\%12\relax}%
\providecommand \@@startlink[1]{}%
\providecommand \@@endlink[0]{}%
\providecommand \url  [0]{\begingroup\@sanitize@url \@url }%
\providecommand \@url [1]{\endgroup\@href {#1}{\urlprefix }}%
\providecommand \urlprefix  [0]{URL }%
\providecommand \Eprint [0]{\href }%
\providecommand \doibase [0]{https://doi.org/}%
\providecommand \selectlanguage [0]{\@gobble}%
\providecommand \bibinfo  [0]{\@secondoftwo}%
\providecommand \bibfield  [0]{\@secondoftwo}%
\providecommand \translation [1]{[#1]}%
\providecommand \BibitemOpen [0]{}%
\providecommand \bibitemStop [0]{}%
\providecommand \bibitemNoStop [0]{.\EOS\space}%
\providecommand \EOS [0]{\spacefactor3000\relax}%
\providecommand \BibitemShut  [1]{\csname bibitem#1\endcsname}%
\let\auto@bib@innerbib\@empty
\bibitem [{\citenamefont {Boyd}(2008)}]{Boyd2008}%
  \BibitemOpen
  \bibfield  {author} {\bibinfo {author} {\bibfnamefont {R.~W.}\ \bibnamefont
  {Boyd}},\ }\href {https://doi.org/10.1016/B978-0-12-369470-6.00001-0} {\emph
  {\bibinfo {title} {{Nonlinear Optics}}}},\ \bibinfo {edition} {3rd}\ ed.\
  (\bibinfo  {publisher} {Elsevier},\ \bibinfo {address} {Amsterdam},\ \bibinfo
  {year} {2008})\BibitemShut {NoStop}%
\bibitem [{\citenamefont {Keller}\ \emph {et~al.}(1992)\citenamefont {Keller},
  \citenamefont {Miller}, \citenamefont {Boyd}, \citenamefont {Chiu},
  \citenamefont {Ferguson},\ and\ \citenamefont {Asom}}]{Keller:92}%
  \BibitemOpen
  \bibfield  {author} {\bibinfo {author} {\bibfnamefont {U.}~\bibnamefont
  {Keller}}, \bibinfo {author} {\bibfnamefont {D.~A.~B.}\ \bibnamefont
  {Miller}}, \bibinfo {author} {\bibfnamefont {G.~D.}\ \bibnamefont {Boyd}},
  \bibinfo {author} {\bibfnamefont {T.~H.}\ \bibnamefont {Chiu}}, \bibinfo
  {author} {\bibfnamefont {J.~F.}\ \bibnamefont {Ferguson}},\ and\ \bibinfo
  {author} {\bibfnamefont {M.~T.}\ \bibnamefont {Asom}},\ }\bibfield  {title}
  {\bibinfo {title} {Solid-state low-loss intracavity saturable absorber for
  nd:ylf lasers: an antiresonant semiconductor fabry--perot saturable
  absorber},\ }\href {https://doi.org/10.1364/OL.17.000505} {\bibfield
  {journal} {\bibinfo  {journal} {Opt. Lett.}\ }\textbf {\bibinfo {volume}
  {17}},\ \bibinfo {pages} {505} (\bibinfo {year} {1992})}\BibitemShut
  {NoStop}%
\bibitem [{\citenamefont {Willenberg}\ \emph {et~al.}(2019)\citenamefont
  {Willenberg}, \citenamefont {Maurer}, \citenamefont {Mayer},\ and\
  \citenamefont {Keller}}]{Willenberg2019}%
  \BibitemOpen
  \bibfield  {author} {\bibinfo {author} {\bibfnamefont {B.}~\bibnamefont
  {Willenberg}}, \bibinfo {author} {\bibfnamefont {J.}~\bibnamefont {Maurer}},
  \bibinfo {author} {\bibfnamefont {B.~W.}\ \bibnamefont {Mayer}},\ and\
  \bibinfo {author} {\bibfnamefont {U.}~\bibnamefont {Keller}},\ }\bibfield
  {title} {\bibinfo {title} {{Sub-cycle time resolution of multi-photon
  momentum transfer in strong-field ionization}},\ }\href
  {https://doi.org/10.1038/s41467-019-13409-6} {\bibfield  {journal} {\bibinfo
  {journal} {Nature Communications}\ }\textbf {\bibinfo {volume} {10}},\
  \bibinfo {pages} {1} (\bibinfo {year} {2019})},\ \Eprint
  {https://arxiv.org/abs/1905.09546} {arXiv:1905.09546} \BibitemShut {NoStop}%
\bibitem [{\citenamefont {Tredicucci}\ \emph {et~al.}(1996)\citenamefont
  {Tredicucci}, \citenamefont {Chen}, \citenamefont {Pellegrini}, \citenamefont
  {B\"orger},\ and\ \citenamefont {Bassani}}]{TrediBista}%
  \BibitemOpen
  \bibfield  {author} {\bibinfo {author} {\bibfnamefont {A.}~\bibnamefont
  {Tredicucci}}, \bibinfo {author} {\bibfnamefont {Y.}~\bibnamefont {Chen}},
  \bibinfo {author} {\bibfnamefont {V.}~\bibnamefont {Pellegrini}}, \bibinfo
  {author} {\bibfnamefont {M.}~\bibnamefont {B\"orger}},\ and\ \bibinfo
  {author} {\bibfnamefont {F.}~\bibnamefont {Bassani}},\ }\bibfield  {title}
  {\bibinfo {title} {Optical bistability of semiconductor microcavities in the
  strong-coupling regime},\ }\href {https://doi.org/10.1103/PhysRevA.54.3493}
  {\bibfield  {journal} {\bibinfo  {journal} {Phys. Rev. A}\ }\textbf {\bibinfo
  {volume} {54}},\ \bibinfo {pages} {3493} (\bibinfo {year}
  {1996})}\BibitemShut {NoStop}%
\bibitem [{\citenamefont {Baas}\ \emph {et~al.}(2004)\citenamefont {Baas},
  \citenamefont {Karr}, \citenamefont {Eleuch},\ and\ \citenamefont
  {Giacobino}}]{giacobino2004}%
  \BibitemOpen
  \bibfield  {author} {\bibinfo {author} {\bibfnamefont {A.}~\bibnamefont
  {Baas}}, \bibinfo {author} {\bibfnamefont {J.~P.}\ \bibnamefont {Karr}},
  \bibinfo {author} {\bibfnamefont {H.}~\bibnamefont {Eleuch}},\ and\ \bibinfo
  {author} {\bibfnamefont {E.}~\bibnamefont {Giacobino}},\ }\bibfield  {title}
  {\bibinfo {title} {Optical bistability in semiconductor microcavities},\
  }\href {https://doi.org/10.1103/PhysRevA.69.023809} {\bibfield  {journal}
  {\bibinfo  {journal} {Phys. Rev. A}\ }\textbf {\bibinfo {volume} {69}},\
  \bibinfo {pages} {023809} (\bibinfo {year} {2004})}\BibitemShut {NoStop}%
\bibitem [{\citenamefont {Helm}(1999)}]{helm1999basic}%
  \BibitemOpen
  \bibfield  {author} {\bibinfo {author} {\bibfnamefont {M.}~\bibnamefont
  {Helm}},\ }\bibfield  {title} {\bibinfo {title} {The basic physics of
  intersubband transitions},\ }in\ \href@noop {} {\emph {\bibinfo {booktitle}
  {Semiconductors and Semimetals}}},\ Vol.~\bibinfo {volume} {62},\ \bibinfo
  {editor} {edited by\ \bibinfo {editor} {\bibfnamefont {H.}~\bibnamefont
  {Liu}}\ and\ \bibinfo {editor} {\bibfnamefont {F.}~\bibnamefont {Capasso}}}\
  (\bibinfo  {publisher} {Elsevier},\ \bibinfo {year} {1999})\ Chap.~\bibinfo
  {chapter} {1}, pp.\ \bibinfo {pages} {1--99}\BibitemShut {NoStop}%
\bibitem [{\citenamefont {Baranov}\ and\ \citenamefont
  {Tournie}(2013)}]{BaranovBook}%
  \BibitemOpen
  \bibfield  {author} {\bibinfo {author} {\bibfnamefont {A.}~\bibnamefont
  {Baranov}}\ and\ \bibinfo {author} {\bibfnamefont {E.}~\bibnamefont
  {Tournie}},\ }\href
  {https://www.sciencedirect.com/book/9780857091215/semiconductor-lasers}
  {\emph {\bibinfo {title} {{Semiconductor lasers : fundamentals and
  applications}}}}\ (\bibinfo  {publisher} {Woodhead Publishing},\ \bibinfo
  {year} {2013})\BibitemShut {NoStop}%
\bibitem [{\citenamefont {Rosencher}\ \emph {et~al.}(2002)\citenamefont
  {Rosencher}, \citenamefont {Vinter}, \citenamefont {Piva},\ and\
  \citenamefont {(Firm)}}]{rosencher2002optoelectronics}%
  \BibitemOpen
  \bibfield  {author} {\bibinfo {author} {\bibfnamefont {E.}~\bibnamefont
  {Rosencher}}, \bibinfo {author} {\bibfnamefont {B.}~\bibnamefont {Vinter}},
  \bibinfo {author} {\bibfnamefont {P.}~\bibnamefont {Piva}},\ and\ \bibinfo
  {author} {\bibfnamefont {K.}~\bibnamefont {(Firm)}},\ }\href
  {https://books.google.fr/books?id=AlL4Oa5jSk8C} {\emph {\bibinfo {title}
  {Optoelectronics}}},\ Knovel Library\ (\bibinfo  {publisher} {Cambridge
  University Press},\ \bibinfo {year} {2002})\BibitemShut {NoStop}%
\bibitem [{\citenamefont {Seilmeier}\ \emph {et~al.}(1987)\citenamefont
  {Seilmeier}, \citenamefont {H\"ubner}, \citenamefont {Abstreiter},
  \citenamefont {Weimann},\ and\ \citenamefont
  {Schlapp}}]{Seilmeier_IntersubbandBleaching}%
  \BibitemOpen
  \bibfield  {author} {\bibinfo {author} {\bibfnamefont {A.}~\bibnamefont
  {Seilmeier}}, \bibinfo {author} {\bibfnamefont {H.-J.}\ \bibnamefont
  {H\"ubner}}, \bibinfo {author} {\bibfnamefont {G.}~\bibnamefont
  {Abstreiter}}, \bibinfo {author} {\bibfnamefont {G.}~\bibnamefont
  {Weimann}},\ and\ \bibinfo {author} {\bibfnamefont {W.}~\bibnamefont
  {Schlapp}},\ }\bibfield  {title} {\bibinfo {title} {Intersubband relaxation
  in
  gaas-${\mathrm{al}}_{\mathrm{x}}$${\mathrm{ga}}_{1\mathrm{\ensuremath{-}}\mathrm{x}}$as
  quantum well structures observed directly by an infrared bleaching
  technique},\ }\href {https://doi.org/10.1103/PhysRevLett.59.1345} {\bibfield
  {journal} {\bibinfo  {journal} {Phys. Rev. Lett.}\ }\textbf {\bibinfo
  {volume} {59}},\ \bibinfo {pages} {1345} (\bibinfo {year}
  {1987})}\BibitemShut {NoStop}%
\bibitem [{\citenamefont {Julien}\ \emph {et~al.}(1988)\citenamefont {Julien},
  \citenamefont {Lourtioz}, \citenamefont {Herschkorn}, \citenamefont
  {Delacourt}, \citenamefont {Pocholle}, \citenamefont {Papuchon},
  \citenamefont {Planel},\ and\ \citenamefont {Le~Roux}}]{Julien-Saturation}%
  \BibitemOpen
  \bibfield  {author} {\bibinfo {author} {\bibfnamefont {F.~H.}\ \bibnamefont
  {Julien}}, \bibinfo {author} {\bibfnamefont {J.}~\bibnamefont {Lourtioz}},
  \bibinfo {author} {\bibfnamefont {N.}~\bibnamefont {Herschkorn}}, \bibinfo
  {author} {\bibfnamefont {D.}~\bibnamefont {Delacourt}}, \bibinfo {author}
  {\bibfnamefont {J.~P.}\ \bibnamefont {Pocholle}}, \bibinfo {author}
  {\bibfnamefont {M.}~\bibnamefont {Papuchon}}, \bibinfo {author}
  {\bibfnamefont {R.}~\bibnamefont {Planel}},\ and\ \bibinfo {author}
  {\bibfnamefont {G.}~\bibnamefont {Le~Roux}},\ }\bibfield  {title} {\bibinfo
  {title} {Optical saturation of intersubband absorption in
  {GaAs-Al$_x$Ga$_{1-x}$As} quantum wells},\ }\href
  {https://doi.org/10.1063/1.100386} {\bibfield  {journal} {\bibinfo  {journal}
  {Applied Physics Letters}\ }\textbf {\bibinfo {volume} {53}},\ \bibinfo
  {pages} {116} (\bibinfo {year} {1988})}\BibitemShut {NoStop}%
\bibitem [{\citenamefont {Vodopyanov}\ \emph {et~al.}(1997)\citenamefont
  {Vodopyanov}, \citenamefont {Chazapis}, \citenamefont {Phillips},
  \citenamefont {Sung},\ and\ \citenamefont {Harris}}]{Vodopyanov_1997}%
  \BibitemOpen
  \bibfield  {author} {\bibinfo {author} {\bibfnamefont {K.~L.}\ \bibnamefont
  {Vodopyanov}}, \bibinfo {author} {\bibfnamefont {V.}~\bibnamefont
  {Chazapis}}, \bibinfo {author} {\bibfnamefont {C.~C.}\ \bibnamefont
  {Phillips}}, \bibinfo {author} {\bibfnamefont {B.}~\bibnamefont {Sung}},\
  and\ \bibinfo {author} {\bibfnamefont {J.~S.}\ \bibnamefont {Harris}},\
  }\bibfield  {title} {\bibinfo {title} {Intersubband absorption saturation
  study of narrow {III-V} multiple quantum wells in the {$\lambda$} = 2.8-9
  {$\mu$}m spectral range},\ }\href
  {https://doi.org/10.1088/0268-1242/12/6/011} {\bibfield  {journal} {\bibinfo
  {journal} {Semiconductor Science and Technology}\ }\textbf {\bibinfo {volume}
  {12}},\ \bibinfo {pages} {708} (\bibinfo {year} {1997})}\BibitemShut
  {NoStop}%
\bibitem [{\citenamefont {Dini}\ \emph {et~al.}(2003)\citenamefont {Dini},
  \citenamefont {K\"ohler}, \citenamefont {Tredicucci}, \citenamefont
  {Biasiol},\ and\ \citenamefont {Sorba}}]{Dini2003}%
  \BibitemOpen
  \bibfield  {author} {\bibinfo {author} {\bibfnamefont {D.}~\bibnamefont
  {Dini}}, \bibinfo {author} {\bibfnamefont {R.}~\bibnamefont {K\"ohler}},
  \bibinfo {author} {\bibfnamefont {A.}~\bibnamefont {Tredicucci}}, \bibinfo
  {author} {\bibfnamefont {G.}~\bibnamefont {Biasiol}},\ and\ \bibinfo {author}
  {\bibfnamefont {L.}~\bibnamefont {Sorba}},\ }\bibfield  {title} {\bibinfo
  {title} {Microcavity polariton splitting of intersubband transitions},\
  }\href {https://doi.org/10.1103/PhysRevLett.90.116401} {\bibfield  {journal}
  {\bibinfo  {journal} {Phys. Rev. Lett.}\ }\textbf {\bibinfo {volume} {90}},\
  \bibinfo {pages} {116401} (\bibinfo {year} {2003})}\BibitemShut {NoStop}%
\bibitem [{\citenamefont {De~Liberato}\ and\ \citenamefont
  {Ciuti}(2009)}]{DeLiberato_StimulatedScattering}%
  \BibitemOpen
  \bibfield  {author} {\bibinfo {author} {\bibfnamefont {S.}~\bibnamefont
  {De~Liberato}}\ and\ \bibinfo {author} {\bibfnamefont {C.}~\bibnamefont
  {Ciuti}},\ }\bibfield  {title} {\bibinfo {title} {Stimulated scattering and
  lasing of intersubband cavity polaritons},\ }\href
  {https://doi.org/10.1103/PhysRevLett.102.136403} {\bibfield  {journal}
  {\bibinfo  {journal} {Phys. Rev. Lett.}\ }\textbf {\bibinfo {volume} {102}},\
  \bibinfo {pages} {136403} (\bibinfo {year} {2009})}\BibitemShut {NoStop}%
\bibitem [{\citenamefont {Delteil}\ \emph {et~al.}(2011)\citenamefont
  {Delteil}, \citenamefont {Vasanelli}, \citenamefont {Jouy}, \citenamefont
  {Barate}, \citenamefont {Moreno}, \citenamefont {Teissier}, \citenamefont
  {Baranov},\ and\ \citenamefont {Sirtori}}]{Delteil_OpticalPhononScattering}%
  \BibitemOpen
  \bibfield  {author} {\bibinfo {author} {\bibfnamefont {A.}~\bibnamefont
  {Delteil}}, \bibinfo {author} {\bibfnamefont {A.}~\bibnamefont {Vasanelli}},
  \bibinfo {author} {\bibfnamefont {P.}~\bibnamefont {Jouy}}, \bibinfo {author}
  {\bibfnamefont {D.}~\bibnamefont {Barate}}, \bibinfo {author} {\bibfnamefont
  {J.~C.}\ \bibnamefont {Moreno}}, \bibinfo {author} {\bibfnamefont
  {R.}~\bibnamefont {Teissier}}, \bibinfo {author} {\bibfnamefont {A.~N.}\
  \bibnamefont {Baranov}},\ and\ \bibinfo {author} {\bibfnamefont
  {C.}~\bibnamefont {Sirtori}},\ }\bibfield  {title} {\bibinfo {title} {Optical
  phonon scattering of cavity polaritons in an electroluminescent device},\
  }\href {https://doi.org/10.1103/PhysRevB.83.081404} {\bibfield  {journal}
  {\bibinfo  {journal} {Phys. Rev. B}\ }\textbf {\bibinfo {volume} {83}},\
  \bibinfo {pages} {081404} (\bibinfo {year} {2011})}\BibitemShut {NoStop}%
\bibitem [{\citenamefont {Zanotto}\ \emph {et~al.}(2015)\citenamefont
  {Zanotto}, \citenamefont {Bianco}, \citenamefont {Sorba}, \citenamefont
  {Biasiol},\ and\ \citenamefont {Tredicucci}}]{Zanotto-Saturation}%
  \BibitemOpen
  \bibfield  {author} {\bibinfo {author} {\bibfnamefont {S.}~\bibnamefont
  {Zanotto}}, \bibinfo {author} {\bibfnamefont {F.}~\bibnamefont {Bianco}},
  \bibinfo {author} {\bibfnamefont {L.}~\bibnamefont {Sorba}}, \bibinfo
  {author} {\bibfnamefont {G.}~\bibnamefont {Biasiol}},\ and\ \bibinfo {author}
  {\bibfnamefont {A.}~\bibnamefont {Tredicucci}},\ }\bibfield  {title}
  {\bibinfo {title} {Saturation and bistability of defect-mode intersubband
  polaritons},\ }\href {https://doi.org/10.1103/PhysRevB.91.085308} {\bibfield
  {journal} {\bibinfo  {journal} {Phys. Rev. B}\ }\textbf {\bibinfo {volume}
  {91}},\ \bibinfo {pages} {085308} (\bibinfo {year} {2015})}\BibitemShut
  {NoStop}%
\bibitem [{\citenamefont {Colombelli}\ and\ \citenamefont
  {Manceau}(2015)}]{Colombelli2015}%
  \BibitemOpen
  \bibfield  {author} {\bibinfo {author} {\bibfnamefont {R.}~\bibnamefont
  {Colombelli}}\ and\ \bibinfo {author} {\bibfnamefont {J.-M.}\ \bibnamefont
  {Manceau}},\ }\bibfield  {title} {\bibinfo {title} {Perspectives for
  intersubband polariton lasers},\ }\href
  {https://doi.org/10.1103/PhysRevX.5.011031} {\bibfield  {journal} {\bibinfo
  {journal} {Phys. Rev. X}\ }\textbf {\bibinfo {volume} {5}},\ \bibinfo {pages}
  {011031} (\bibinfo {year} {2015})}\BibitemShut {NoStop}%
\bibitem [{\citenamefont {Manceau}\ \emph {et~al.}(2018)\citenamefont
  {Manceau}, \citenamefont {Tran}, \citenamefont {Biasiol}, \citenamefont
  {Laurent}, \citenamefont {Sagnes}, \citenamefont {Beaudoin}, \citenamefont
  {De~Liberato}, \citenamefont {Carusotto},\ and\ \citenamefont
  {Colombelli}}]{Manceau2018}%
  \BibitemOpen
  \bibfield  {author} {\bibinfo {author} {\bibfnamefont {J.-M.}\ \bibnamefont
  {Manceau}}, \bibinfo {author} {\bibfnamefont {N.-L.}\ \bibnamefont {Tran}},
  \bibinfo {author} {\bibfnamefont {G.}~\bibnamefont {Biasiol}}, \bibinfo
  {author} {\bibfnamefont {T.}~\bibnamefont {Laurent}}, \bibinfo {author}
  {\bibfnamefont {I.}~\bibnamefont {Sagnes}}, \bibinfo {author} {\bibfnamefont
  {G.}~\bibnamefont {Beaudoin}}, \bibinfo {author} {\bibfnamefont
  {S.}~\bibnamefont {De~Liberato}}, \bibinfo {author} {\bibfnamefont
  {I.}~\bibnamefont {Carusotto}},\ and\ \bibinfo {author} {\bibfnamefont
  {R.}~\bibnamefont {Colombelli}},\ }\bibfield  {title} {\bibinfo {title}
  {Resonant intersubband polariton-lo phonon scattering in an optically pumped
  polaritonic device},\ }\href@noop {} {\bibfield  {journal} {\bibinfo
  {journal} {Applied Physics Letters}\ }\textbf {\bibinfo {volume} {112}},\
  \bibinfo {pages} {191106} (\bibinfo {year} {2018})}\BibitemShut {NoStop}%
\bibitem [{\citenamefont {Todorov}\ \emph {et~al.}(2010)\citenamefont
  {Todorov}, \citenamefont {Andrews}, \citenamefont {Colombelli}, \citenamefont
  {De~Liberato}, \citenamefont {Ciuti}, \citenamefont {Klang}, \citenamefont
  {Strasser},\ and\ \citenamefont {Sirtori}}]{todorov2010ultrastrong}%
  \BibitemOpen
  \bibfield  {author} {\bibinfo {author} {\bibfnamefont {Y.}~\bibnamefont
  {Todorov}}, \bibinfo {author} {\bibfnamefont {A.~M.}\ \bibnamefont
  {Andrews}}, \bibinfo {author} {\bibfnamefont {R.}~\bibnamefont {Colombelli}},
  \bibinfo {author} {\bibfnamefont {S.}~\bibnamefont {De~Liberato}}, \bibinfo
  {author} {\bibfnamefont {C.}~\bibnamefont {Ciuti}}, \bibinfo {author}
  {\bibfnamefont {P.}~\bibnamefont {Klang}}, \bibinfo {author} {\bibfnamefont
  {G.}~\bibnamefont {Strasser}},\ and\ \bibinfo {author} {\bibfnamefont
  {C.}~\bibnamefont {Sirtori}},\ }\bibfield  {title} {\bibinfo {title}
  {Ultrastrong light-matter coupling regime with polariton dots},\ }\href
  {https://doi.org/10.1103/PhysRevLett.105.196402} {\bibfield  {journal}
  {\bibinfo  {journal} {Physical Review Letters}\ }\textbf {\bibinfo {volume}
  {105}},\ \bibinfo {pages} {196402} (\bibinfo {year} {2010})}\BibitemShut
  {NoStop}%
\bibitem [{\citenamefont {Scalari}\ \emph {et~al.}(2012)\citenamefont
  {Scalari}, \citenamefont {Maissen}, \citenamefont {Turcinkova}, \citenamefont
  {Hagenmuller}, \citenamefont {{De Liberato}}, \citenamefont {Ciuti},
  \citenamefont {Reichl}, \citenamefont {Schuh}, \citenamefont {Wegscheider},
  \citenamefont {Beck},\ and\ \citenamefont {Faist}}]{Scalari2012}%
  \BibitemOpen
  \bibfield  {author} {\bibinfo {author} {\bibfnamefont {G.}~\bibnamefont
  {Scalari}}, \bibinfo {author} {\bibfnamefont {C.}~\bibnamefont {Maissen}},
  \bibinfo {author} {\bibfnamefont {D.}~\bibnamefont {Turcinkova}}, \bibinfo
  {author} {\bibfnamefont {D.}~\bibnamefont {Hagenmuller}}, \bibinfo {author}
  {\bibfnamefont {S.}~\bibnamefont {{De Liberato}}}, \bibinfo {author}
  {\bibfnamefont {C.}~\bibnamefont {Ciuti}}, \bibinfo {author} {\bibfnamefont
  {C.}~\bibnamefont {Reichl}}, \bibinfo {author} {\bibfnamefont
  {D.}~\bibnamefont {Schuh}}, \bibinfo {author} {\bibfnamefont
  {W.}~\bibnamefont {Wegscheider}}, \bibinfo {author} {\bibfnamefont
  {M.}~\bibnamefont {Beck}},\ and\ \bibinfo {author} {\bibfnamefont
  {J.}~\bibnamefont {Faist}},\ }\bibfield  {title} {\bibinfo {title}
  {{Ultrastrong Coupling of the Cyclotron Transition of a 2D Electron Gas to a
  THz Metamaterial}},\ }\href {https://doi.org/10.1126/science.1216022}
  {\bibfield  {journal} {\bibinfo  {journal} {Science}\ }\textbf {\bibinfo
  {volume} {335}},\ \bibinfo {pages} {1323} (\bibinfo {year} {2012})},\ \Eprint
  {https://arxiv.org/abs/1111.2486} {arXiv:1111.2486} \BibitemShut {NoStop}%
\bibitem [{\citenamefont {Delteil}\ \emph {et~al.}(2012)\citenamefont
  {Delteil}, \citenamefont {Vasanelli}, \citenamefont {Todorov}, \citenamefont
  {{Feuillet Palma}}, \citenamefont {{Renaudat St-Jean}}, \citenamefont
  {Beaudoin}, \citenamefont {Sagnes},\ and\ \citenamefont
  {Sirtori}}]{Delteil2012}%
  \BibitemOpen
  \bibfield  {author} {\bibinfo {author} {\bibfnamefont {A.}~\bibnamefont
  {Delteil}}, \bibinfo {author} {\bibfnamefont {A.}~\bibnamefont {Vasanelli}},
  \bibinfo {author} {\bibfnamefont {Y.}~\bibnamefont {Todorov}}, \bibinfo
  {author} {\bibfnamefont {C.}~\bibnamefont {{Feuillet Palma}}}, \bibinfo
  {author} {\bibfnamefont {M.}~\bibnamefont {{Renaudat St-Jean}}}, \bibinfo
  {author} {\bibfnamefont {G.}~\bibnamefont {Beaudoin}}, \bibinfo {author}
  {\bibfnamefont {I.}~\bibnamefont {Sagnes}},\ and\ \bibinfo {author}
  {\bibfnamefont {C.}~\bibnamefont {Sirtori}},\ }\bibfield  {title} {\bibinfo
  {title} {{Charge-Induced Coherence between Intersubband Plasmons in a Quantum
  Structure}},\ }\href {https://doi.org/10.1103/PhysRevLett.109.246808}
  {\bibfield  {journal} {\bibinfo  {journal} {Phys. Rev. Lett.}\ }\textbf
  {\bibinfo {volume} {109}},\ \bibinfo {pages} {246808} (\bibinfo {year}
  {2012})},\ \Eprint {https://arxiv.org/abs/1212.4422} {arXiv:1212.4422}
  \BibitemShut {NoStop}%
\bibitem [{\citenamefont {Malerba}\ \emph {et~al.}(2016)\citenamefont
  {Malerba}, \citenamefont {Ongarello}, \citenamefont {Paulillo}, \citenamefont
  {Manceau}, \citenamefont {Beaudoin}, \citenamefont {Sagnes}, \citenamefont
  {De~Angelis},\ and\ \citenamefont {Colombelli}}]{Malerba2017}%
  \BibitemOpen
  \bibfield  {author} {\bibinfo {author} {\bibfnamefont {M.}~\bibnamefont
  {Malerba}}, \bibinfo {author} {\bibfnamefont {T.}~\bibnamefont {Ongarello}},
  \bibinfo {author} {\bibfnamefont {B.}~\bibnamefont {Paulillo}}, \bibinfo
  {author} {\bibfnamefont {J.-M.}\ \bibnamefont {Manceau}}, \bibinfo {author}
  {\bibfnamefont {G.}~\bibnamefont {Beaudoin}}, \bibinfo {author}
  {\bibfnamefont {I.}~\bibnamefont {Sagnes}}, \bibinfo {author} {\bibfnamefont
  {F.}~\bibnamefont {De~Angelis}},\ and\ \bibinfo {author} {\bibfnamefont
  {R.}~\bibnamefont {Colombelli}},\ }\bibfield  {title} {\bibinfo {title}
  {Towards strong light-matter coupling at the single-resonator level with
  sub-wavelength mid-infrared nano-antennas},\ }\href
  {https://doi.org/10.1063/1.4958330} {\bibfield  {journal} {\bibinfo
  {journal} {Applied Physics Letters}\ }\textbf {\bibinfo {volume} {109}},\
  \bibinfo {pages} {021111} (\bibinfo {year} {2016})},\ \Eprint
  {https://arxiv.org/abs/https://doi.org/10.1063/1.4958330}
  {https://doi.org/10.1063/1.4958330} \BibitemShut {NoStop}%
\bibitem [{\citenamefont {Keller}\ \emph {et~al.}(2017)\citenamefont {Keller},
  \citenamefont {Scalari}, \citenamefont {Cibella}, \citenamefont {Maissen},
  \citenamefont {Appugliese}, \citenamefont {Giovine}, \citenamefont {Leoni},
  \citenamefont {Beck},\ and\ \citenamefont {Faist}}]{Keller2017}%
  \BibitemOpen
  \bibfield  {author} {\bibinfo {author} {\bibfnamefont {J.}~\bibnamefont
  {Keller}}, \bibinfo {author} {\bibfnamefont {G.}~\bibnamefont {Scalari}},
  \bibinfo {author} {\bibfnamefont {S.}~\bibnamefont {Cibella}}, \bibinfo
  {author} {\bibfnamefont {C.}~\bibnamefont {Maissen}}, \bibinfo {author}
  {\bibfnamefont {F.}~\bibnamefont {Appugliese}}, \bibinfo {author}
  {\bibfnamefont {E.}~\bibnamefont {Giovine}}, \bibinfo {author} {\bibfnamefont
  {R.}~\bibnamefont {Leoni}}, \bibinfo {author} {\bibfnamefont
  {M.}~\bibnamefont {Beck}},\ and\ \bibinfo {author} {\bibfnamefont
  {J.}~\bibnamefont {Faist}},\ }\bibfield  {title} {\bibinfo {title}
  {Few-electron ultrastrong light-matter coupling at 300 ghz with nanogap
  hybrid lc microcavities},\ }\href
  {https://doi.org/10.1021/acs.nanolett.7b03228} {\bibfield  {journal}
  {\bibinfo  {journal} {Nano Letters}\ }\textbf {\bibinfo {volume} {17}},\
  \bibinfo {pages} {7410} (\bibinfo {year} {2017})},\ \bibinfo {note} {pMID:
  29172537},\ \Eprint
  {https://arxiv.org/abs/https://doi.org/10.1021/acs.nanolett.7b03228}
  {https://doi.org/10.1021/acs.nanolett.7b03228} \BibitemShut {NoStop}%
\bibitem [{\citenamefont {Jeannin}\ \emph {et~al.}(2019)\citenamefont
  {Jeannin}, \citenamefont {Mariotti~Nesurini}, \citenamefont {Suffit},
  \citenamefont {Gacemi}, \citenamefont {Vasanelli}, \citenamefont {Li},
  \citenamefont {Davies}, \citenamefont {Linfield}, \citenamefont {Sirtori},\
  and\ \citenamefont {Todorov}}]{Jeannin_Ultrastrong}%
  \BibitemOpen
  \bibfield  {author} {\bibinfo {author} {\bibfnamefont {M.}~\bibnamefont
  {Jeannin}}, \bibinfo {author} {\bibfnamefont {G.}~\bibnamefont
  {Mariotti~Nesurini}}, \bibinfo {author} {\bibfnamefont {S.}~\bibnamefont
  {Suffit}}, \bibinfo {author} {\bibfnamefont {D.}~\bibnamefont {Gacemi}},
  \bibinfo {author} {\bibfnamefont {A.}~\bibnamefont {Vasanelli}}, \bibinfo
  {author} {\bibfnamefont {L.~H.}\ \bibnamefont {Li}}, \bibinfo {author}
  {\bibfnamefont {A.~G.}\ \bibnamefont {Davies}}, \bibinfo {author}
  {\bibfnamefont {E.~H.}\ \bibnamefont {Linfield}}, \bibinfo {author}
  {\bibfnamefont {C.}~\bibnamefont {Sirtori}},\ and\ \bibinfo {author}
  {\bibfnamefont {Y.}~\bibnamefont {Todorov}},\ }\bibfield  {title} {\bibinfo
  {title} {Ultra-strong light-matter coupling in deeply subwavelength thz lc
  resonators},\ }\href {https://doi.org/10.1021/acsphotonics.8b01778}
  {\bibfield  {journal} {\bibinfo  {journal} {ACS Photonics}\ }\textbf
  {\bibinfo {volume} {6}},\ \bibinfo {pages} {1207} (\bibinfo {year}
  {2019})}\BibitemShut {NoStop}%
\bibitem [{\citenamefont {Ciuti}\ \emph {et~al.}(2005)\citenamefont {Ciuti},
  \citenamefont {Bastard},\ and\ \citenamefont {Carusotto}}]{ciuti2005quantum}%
  \BibitemOpen
  \bibfield  {author} {\bibinfo {author} {\bibfnamefont {C.}~\bibnamefont
  {Ciuti}}, \bibinfo {author} {\bibfnamefont {G.}~\bibnamefont {Bastard}},\
  and\ \bibinfo {author} {\bibfnamefont {I.}~\bibnamefont {Carusotto}},\
  }\bibfield  {title} {\bibinfo {title} {Quantum vacuum properties of the
  intersubband cavity polariton field},\ }\href@noop {} {\bibfield  {journal}
  {\bibinfo  {journal} {Physical Review B}\ }\textbf {\bibinfo {volume} {72}},\
  \bibinfo {pages} {115303} (\bibinfo {year} {2005})}\BibitemShut {NoStop}%
\bibitem [{\citenamefont {G{\"{u}}nter}\ \emph {et~al.}(2009)\citenamefont
  {G{\"{u}}nter}, \citenamefont {Anappara}, \citenamefont {Hees}, \citenamefont
  {Sell}, \citenamefont {Biasiol}, \citenamefont {Sorba}, \citenamefont {{De
  Liberato}}, \citenamefont {Ciuti}, \citenamefont {Tredicucci}, \citenamefont
  {Leitenstorfer},\ and\ \citenamefont {Huber}}]{Gunter2009}%
  \BibitemOpen
  \bibfield  {author} {\bibinfo {author} {\bibfnamefont {G.}~\bibnamefont
  {G{\"{u}}nter}}, \bibinfo {author} {\bibfnamefont {A.~A.}\ \bibnamefont
  {Anappara}}, \bibinfo {author} {\bibfnamefont {J.}~\bibnamefont {Hees}},
  \bibinfo {author} {\bibfnamefont {A.}~\bibnamefont {Sell}}, \bibinfo {author}
  {\bibfnamefont {G.}~\bibnamefont {Biasiol}}, \bibinfo {author} {\bibfnamefont
  {L.}~\bibnamefont {Sorba}}, \bibinfo {author} {\bibfnamefont
  {S.}~\bibnamefont {{De Liberato}}}, \bibinfo {author} {\bibfnamefont
  {C.}~\bibnamefont {Ciuti}}, \bibinfo {author} {\bibfnamefont
  {A.}~\bibnamefont {Tredicucci}}, \bibinfo {author} {\bibfnamefont
  {A.}~\bibnamefont {Leitenstorfer}},\ and\ \bibinfo {author} {\bibfnamefont
  {R.}~\bibnamefont {Huber}},\ }\bibfield  {title} {\bibinfo {title}
  {{Sub-cycle switch-on of ultrastrong light--matter interaction}},\ }\href
  {https://doi.org/10.1038/nature07838} {\bibfield  {journal} {\bibinfo
  {journal} {Nature}\ }\textbf {\bibinfo {volume} {458}},\ \bibinfo {pages}
  {178} (\bibinfo {year} {2009})}\BibitemShut {NoStop}%
\bibitem [{\citenamefont {Lee}\ \emph {et~al.}(2014)\citenamefont {Lee},
  \citenamefont {Jung}, \citenamefont {Chen}, \citenamefont {Lu}, \citenamefont
  {Demmerle}, \citenamefont {Boehm}, \citenamefont {Amann}, \citenamefont
  {AlÃ¹},\ and\ \citenamefont {Belkin}}]{Lee2014}%
  \BibitemOpen
  \bibfield  {author} {\bibinfo {author} {\bibfnamefont {J.}~\bibnamefont
  {Lee}}, \bibinfo {author} {\bibfnamefont {S.}~\bibnamefont {Jung}}, \bibinfo
  {author} {\bibfnamefont {P.-Y.}\ \bibnamefont {Chen}}, \bibinfo {author}
  {\bibfnamefont {F.}~\bibnamefont {Lu}}, \bibinfo {author} {\bibfnamefont
  {F.}~\bibnamefont {Demmerle}}, \bibinfo {author} {\bibfnamefont
  {G.}~\bibnamefont {Boehm}}, \bibinfo {author} {\bibfnamefont {M.-C.}\
  \bibnamefont {Amann}}, \bibinfo {author} {\bibfnamefont {A.}~\bibnamefont
  {AlÃ¹}},\ and\ \bibinfo {author} {\bibfnamefont {M.~A.}\ \bibnamefont
  {Belkin}},\ }\bibfield  {title} {\bibinfo {title} {Ultrafast electrically
  tunable polaritonic metasurfaces},\ }\href
  {https://doi.org/10.1002/adom.201400185} {\bibfield  {journal} {\bibinfo
  {journal} {Advanced Optical Materials}\ }\textbf {\bibinfo {volume} {2}},\
  \bibinfo {pages} {1057} (\bibinfo {year} {2014})}\BibitemShut {NoStop}%
\bibitem [{\citenamefont {Pirotta}\ \emph {et~al.}(2020)\citenamefont
  {Pirotta}, \citenamefont {Tran}, \citenamefont {Biasiol}, \citenamefont
  {Jollivet}, \citenamefont {Crozat}, \citenamefont {Manceau}, \citenamefont
  {Bousseksou},\ and\ \citenamefont {Colombelli}}]{pirotta2020ultrafast}%
  \BibitemOpen
  \bibfield  {author} {\bibinfo {author} {\bibfnamefont {S.}~\bibnamefont
  {Pirotta}}, \bibinfo {author} {\bibfnamefont {N.-L.}\ \bibnamefont {Tran}},
  \bibinfo {author} {\bibfnamefont {G.}~\bibnamefont {Biasiol}}, \bibinfo
  {author} {\bibfnamefont {A.}~\bibnamefont {Jollivet}}, \bibinfo {author}
  {\bibfnamefont {P.}~\bibnamefont {Crozat}}, \bibinfo {author} {\bibfnamefont
  {J.-M.}\ \bibnamefont {Manceau}}, \bibinfo {author} {\bibfnamefont
  {A.}~\bibnamefont {Bousseksou}},\ and\ \bibinfo {author} {\bibfnamefont
  {R.}~\bibnamefont {Colombelli}},\ }\bibfield  {title} {\bibinfo {title}
  {Ultra-fast amplitude modulation of mid-ir free-space beams at
  room-temperature},\ }\href@noop {} {\bibfield  {journal} {\bibinfo  {journal}
  {arXiv}\ ,\ \bibinfo {pages} {2006.12215}} (\bibinfo {year} {2020})},\
  \Eprint {https://arxiv.org/abs/2006.12215} {arXiv:2006.12215
  [physics.optics]} \BibitemShut {NoStop}%
\bibitem [{\citenamefont {Vigneron}\ \emph {et~al.}(2019)\citenamefont
  {Vigneron}, \citenamefont {Pirotta}, \citenamefont {Carusotto}, \citenamefont
  {Tran}, \citenamefont {Biasiol}, \citenamefont {Manceau}, \citenamefont
  {Bousseksou},\ and\ \citenamefont {Colombelli}}]{Vigneron2019}%
  \BibitemOpen
  \bibfield  {author} {\bibinfo {author} {\bibfnamefont {P.-B.}\ \bibnamefont
  {Vigneron}}, \bibinfo {author} {\bibfnamefont {S.}~\bibnamefont {Pirotta}},
  \bibinfo {author} {\bibfnamefont {I.}~\bibnamefont {Carusotto}}, \bibinfo
  {author} {\bibfnamefont {N.-L.}\ \bibnamefont {Tran}}, \bibinfo {author}
  {\bibfnamefont {G.}~\bibnamefont {Biasiol}}, \bibinfo {author} {\bibfnamefont
  {J.-M.}\ \bibnamefont {Manceau}}, \bibinfo {author} {\bibfnamefont
  {A.}~\bibnamefont {Bousseksou}},\ and\ \bibinfo {author} {\bibfnamefont
  {R.}~\bibnamefont {Colombelli}},\ }\bibfield  {title} {\bibinfo {title}
  {Quantum well infrared photo-detectors operating in the strong light-matter
  coupling regime},\ }\href@noop {} {\bibfield  {journal} {\bibinfo  {journal}
  {Applied Physics Letters}\ }\textbf {\bibinfo {volume} {114}},\ \bibinfo
  {pages} {131104} (\bibinfo {year} {2019})}\BibitemShut {NoStop}%
\bibitem [{\citenamefont {Limbacher}\ \emph {et~al.}(2020)\citenamefont
  {Limbacher}, \citenamefont {Kainz}, \citenamefont {Schoenhuber},
  \citenamefont {Wenclawiak}, \citenamefont {Derntl}, \citenamefont {Andrews},
  \citenamefont {Detz}, \citenamefont {Strasser}, \citenamefont {Schwaighofer},
  \citenamefont {Lendl}, \citenamefont {Darmo},\ and\ \citenamefont
  {Unterrainer}}]{Limbacher_RTD_2020}%
  \BibitemOpen
  \bibfield  {author} {\bibinfo {author} {\bibfnamefont {B.}~\bibnamefont
  {Limbacher}}, \bibinfo {author} {\bibfnamefont {M.~A.}\ \bibnamefont
  {Kainz}}, \bibinfo {author} {\bibfnamefont {S.}~\bibnamefont {Schoenhuber}},
  \bibinfo {author} {\bibfnamefont {M.}~\bibnamefont {Wenclawiak}}, \bibinfo
  {author} {\bibfnamefont {C.}~\bibnamefont {Derntl}}, \bibinfo {author}
  {\bibfnamefont {A.~M.}\ \bibnamefont {Andrews}}, \bibinfo {author}
  {\bibfnamefont {H.}~\bibnamefont {Detz}}, \bibinfo {author} {\bibfnamefont
  {G.}~\bibnamefont {Strasser}}, \bibinfo {author} {\bibfnamefont
  {A.}~\bibnamefont {Schwaighofer}}, \bibinfo {author} {\bibfnamefont
  {B.}~\bibnamefont {Lendl}}, \bibinfo {author} {\bibfnamefont
  {J.}~\bibnamefont {Darmo}},\ and\ \bibinfo {author} {\bibfnamefont
  {K.}~\bibnamefont {Unterrainer}},\ }\bibfield  {title} {\bibinfo {title}
  {Resonant tunneling diodes strongly coupled to the cavity field},\ }\href
  {https://doi.org/10.1063/5.0007118} {\bibfield  {journal} {\bibinfo
  {journal} {Applied Physics Letters}\ }\textbf {\bibinfo {volume} {116}},\
  \bibinfo {pages} {221101} (\bibinfo {year} {2020})}\BibitemShut {NoStop}%
\bibitem [{\citenamefont {Alpeggiani}\ and\ \citenamefont
  {Andreani}(2014)}]{alpeggiani2014semiclassical}%
  \BibitemOpen
  \bibfield  {author} {\bibinfo {author} {\bibfnamefont {F.}~\bibnamefont
  {Alpeggiani}}\ and\ \bibinfo {author} {\bibfnamefont {L.~C.}\ \bibnamefont
  {Andreani}},\ }\bibfield  {title} {\bibinfo {title} {Semiclassical theory of
  multisubband plasmons: Nonlocal electrodynamics and radiative effects},\
  }\href {https://doi.org/10.1103/PhysRevB.90.115311} {\bibfield  {journal}
  {\bibinfo  {journal} {Physical Review B}\ }\textbf {\bibinfo {volume} {90}},\
  \bibinfo {pages} {115311} (\bibinfo {year} {2014})}\BibitemShut {NoStop}%
\bibitem [{\citenamefont {Jeannin}\ \emph {et~al.}(2020)\citenamefont
  {Jeannin}, \citenamefont {Bonazzi}, \citenamefont {Gacemi}, \citenamefont
  {Vasanelli}, \citenamefont {Li}, \citenamefont {Davies}, \citenamefont
  {Linfield}, \citenamefont {Sirtori},\ and\ \citenamefont
  {Todorov}}]{Jeannin-abseng}%
  \BibitemOpen
  \bibfield  {author} {\bibinfo {author} {\bibfnamefont {M.}~\bibnamefont
  {Jeannin}}, \bibinfo {author} {\bibfnamefont {T.}~\bibnamefont {Bonazzi}},
  \bibinfo {author} {\bibfnamefont {D.}~\bibnamefont {Gacemi}}, \bibinfo
  {author} {\bibfnamefont {A.}~\bibnamefont {Vasanelli}}, \bibinfo {author}
  {\bibfnamefont {L.}~\bibnamefont {Li}}, \bibinfo {author} {\bibfnamefont
  {A.~G.}\ \bibnamefont {Davies}}, \bibinfo {author} {\bibfnamefont
  {E.}~\bibnamefont {Linfield}}, \bibinfo {author} {\bibfnamefont
  {C.}~\bibnamefont {Sirtori}},\ and\ \bibinfo {author} {\bibfnamefont
  {Y.}~\bibnamefont {Todorov}},\ }\bibfield  {title} {\bibinfo {title}
  {Absorption engineering in an ultrasubwavelength quantum system},\
  }\href@noop {} {\bibfield  {journal} {\bibinfo  {journal} {Nano Letters}\
  }\textbf {\bibinfo {volume} {20}},\ \bibinfo {pages} {4430} (\bibinfo {year}
  {2020})}\BibitemShut {NoStop}%
\bibitem [{\citenamefont {Warburton}\ \emph {et~al.}(1998)\citenamefont
  {Warburton}, \citenamefont {Weilhammer}, \citenamefont {Kotthaus},
  \citenamefont {Thomas},\ and\ \citenamefont {Kroemer}}]{Warburton_1998}%
  \BibitemOpen
  \bibfield  {author} {\bibinfo {author} {\bibfnamefont {R.~J.}\ \bibnamefont
  {Warburton}}, \bibinfo {author} {\bibfnamefont {K.}~\bibnamefont
  {Weilhammer}}, \bibinfo {author} {\bibfnamefont {J.~P.}\ \bibnamefont
  {Kotthaus}}, \bibinfo {author} {\bibfnamefont {M.}~\bibnamefont {Thomas}},\
  and\ \bibinfo {author} {\bibfnamefont {H.}~\bibnamefont {Kroemer}},\
  }\bibfield  {title} {\bibinfo {title} {Influence of collective effects on the
  linewidth of intersubband resonance},\ }\href
  {https://doi.org/10.1103/PhysRevLett.80.2185} {\bibfield  {journal} {\bibinfo
   {journal} {Physical Review Letters}\ }\textbf {\bibinfo {volume} {80}},\
  \bibinfo {pages} {2185â€“2188} (\bibinfo {year} {1998})}\BibitemShut
  {NoStop}%
\bibitem [{\citenamefont {Tran}\ \emph {et~al.}(2019)\citenamefont {Tran},
  \citenamefont {Biasiol}, \citenamefont {Jollivet}, \citenamefont {Bertocci},
  \citenamefont {Julien}, \citenamefont {Manceau},\ and\ \citenamefont
  {Colombelli}}]{growth-interruptions}%
  \BibitemOpen
  \bibfield  {author} {\bibinfo {author} {\bibfnamefont {N.~L.}\ \bibnamefont
  {Tran}}, \bibinfo {author} {\bibfnamefont {G.}~\bibnamefont {Biasiol}},
  \bibinfo {author} {\bibfnamefont {A.}~\bibnamefont {Jollivet}}, \bibinfo
  {author} {\bibfnamefont {A.}~\bibnamefont {Bertocci}}, \bibinfo {author}
  {\bibfnamefont {F.~H.}\ \bibnamefont {Julien}}, \bibinfo {author}
  {\bibfnamefont {J.-M.}\ \bibnamefont {Manceau}},\ and\ \bibinfo {author}
  {\bibfnamefont {R.}~\bibnamefont {Colombelli}},\ }\bibfield  {title}
  {\bibinfo {title} {Evidence of intersubband linewidth narrowing using growth
  interruption technique},\ }\bibfield  {journal} {\bibinfo  {journal}
  {Photonics}\ }\textbf {\bibinfo {volume} {6}},\ \href
  {https://doi.org/10.3390/photonics6020038} {10.3390/photonics6020038}
  (\bibinfo {year} {2019})\BibitemShut {NoStop}%
\bibitem [{\citenamefont {Lee}\ \emph {et~al.}(2016)\citenamefont {Lee},
  \citenamefont {Nookala}, \citenamefont {Gomez-Diaz}, \citenamefont
  {Tymchenko}, \citenamefont {Demmerle}, \citenamefont {Boehm}, \citenamefont
  {Amann}, \citenamefont {AlÃ¹},\ and\ \citenamefont
  {Belkin}}]{second-harmonic-surface-1}%
  \BibitemOpen
  \bibfield  {author} {\bibinfo {author} {\bibfnamefont {J.}~\bibnamefont
  {Lee}}, \bibinfo {author} {\bibfnamefont {N.}~\bibnamefont {Nookala}},
  \bibinfo {author} {\bibfnamefont {J.~S.}\ \bibnamefont {Gomez-Diaz}},
  \bibinfo {author} {\bibfnamefont {M.}~\bibnamefont {Tymchenko}}, \bibinfo
  {author} {\bibfnamefont {F.}~\bibnamefont {Demmerle}}, \bibinfo {author}
  {\bibfnamefont {G.}~\bibnamefont {Boehm}}, \bibinfo {author} {\bibfnamefont
  {M.-C.}\ \bibnamefont {Amann}}, \bibinfo {author} {\bibfnamefont
  {A.}~\bibnamefont {AlÃ¹}},\ and\ \bibinfo {author} {\bibfnamefont {M.~A.}\
  \bibnamefont {Belkin}},\ }\bibfield  {title} {\bibinfo {title} {Ultrathin
  second-harmonic metasurfaces with record-high nonlinear optical response},\
  }\href@noop {} {\bibfield  {journal} {\bibinfo  {journal} {Advanced Optical
  Materials}\ }\textbf {\bibinfo {volume} {4}},\ \bibinfo {pages} {664}
  (\bibinfo {year} {2016})}\BibitemShut {NoStop}%
\bibitem [{\citenamefont {Roland}\ \emph {et~al.}(2020)\citenamefont {Roland},
  \citenamefont {Borne}, \citenamefont {Ravaro}, \citenamefont {Oliveira},
  \citenamefont {Suffit}, \citenamefont {Filloux}, \citenamefont
  {Lema\^{i}tre}, \citenamefont {Favero},\ and\ \citenamefont
  {Leo}}]{YannisChip}%
  \BibitemOpen
  \bibfield  {author} {\bibinfo {author} {\bibfnamefont {I.}~\bibnamefont
  {Roland}}, \bibinfo {author} {\bibfnamefont {A.}~\bibnamefont {Borne}},
  \bibinfo {author} {\bibfnamefont {M.}~\bibnamefont {Ravaro}}, \bibinfo
  {author} {\bibfnamefont {R.~D.}\ \bibnamefont {Oliveira}}, \bibinfo {author}
  {\bibfnamefont {S.}~\bibnamefont {Suffit}}, \bibinfo {author} {\bibfnamefont
  {P.}~\bibnamefont {Filloux}}, \bibinfo {author} {\bibfnamefont
  {A.}~\bibnamefont {Lema\^{i}tre}}, \bibinfo {author} {\bibfnamefont
  {I.}~\bibnamefont {Favero}},\ and\ \bibinfo {author} {\bibfnamefont
  {G.}~\bibnamefont {Leo}},\ }\bibfield  {title} {\bibinfo {title} {Frequency
  doubling and parametric fluorescence in a four-port aluminum gallium arsenide
  photonic chip},\ }\href {https://doi.org/10.1364/OL.392417} {\bibfield
  {journal} {\bibinfo  {journal} {Opt. Lett.}\ }\textbf {\bibinfo {volume}
  {45}},\ \bibinfo {pages} {2878} (\bibinfo {year} {2020})}\BibitemShut
  {NoStop}%
\bibitem [{\citenamefont {Balanis}(2005)}]{balanis-book}%
  \BibitemOpen
  \bibfield  {author} {\bibinfo {author} {\bibfnamefont {C.~A.}\ \bibnamefont
  {Balanis}},\ }\href@noop {} {\emph {\bibinfo {title} {Antenna Theory:
  Analysis and Design}}}\ (\bibinfo  {publisher} {Wiley-Interscience},\
  \bibinfo {address} {USA},\ \bibinfo {year} {2005})\BibitemShut {NoStop}%
\bibitem [{\citenamefont {Hakl}\ \emph {et~al.}(2020)\citenamefont {Hakl},
  \citenamefont {Lin}, \citenamefont {Lepillet}, \citenamefont {Billet},
  \citenamefont {Lampin}, \citenamefont {Pirotta}, \citenamefont {Colombelli},
  \citenamefont {Wan}, \citenamefont {Cao}, \citenamefont {Li}, \citenamefont
  {Peytavit},\ and\ \citenamefont {Barbieri}}]{hakl2020ultrafast}%
  \BibitemOpen
  \bibfield  {author} {\bibinfo {author} {\bibfnamefont {M.}~\bibnamefont
  {Hakl}}, \bibinfo {author} {\bibfnamefont {Q.~Y.}\ \bibnamefont {Lin}},
  \bibinfo {author} {\bibfnamefont {S.}~\bibnamefont {Lepillet}}, \bibinfo
  {author} {\bibfnamefont {M.}~\bibnamefont {Billet}}, \bibinfo {author}
  {\bibfnamefont {J.-F.}\ \bibnamefont {Lampin}}, \bibinfo {author}
  {\bibfnamefont {S.}~\bibnamefont {Pirotta}}, \bibinfo {author} {\bibfnamefont
  {R.}~\bibnamefont {Colombelli}}, \bibinfo {author} {\bibfnamefont {W.~J.}\
  \bibnamefont {Wan}}, \bibinfo {author} {\bibfnamefont {J.~C.}\ \bibnamefont
  {Cao}}, \bibinfo {author} {\bibfnamefont {H.}~\bibnamefont {Li}}, \bibinfo
  {author} {\bibfnamefont {E.}~\bibnamefont {Peytavit}},\ and\ \bibinfo
  {author} {\bibfnamefont {S.}~\bibnamefont {Barbieri}},\ }\bibfield  {title}
  {\bibinfo {title} {Ultra-fast quantum-well infared photodetectors operating
  at 10 {$\mu$}m with flat response up to 70 ghz at room temperature},\
  }\href@noop {} {\bibfield  {journal} {\bibinfo  {journal} {arXiv 2007.00299}\
  } (\bibinfo {year} {2020})}\BibitemShut {NoStop}%
\bibitem [{\citenamefont {Cortese}\ \emph {et~al.}(2020)\citenamefont
  {Cortese}, \citenamefont {Tran}, \citenamefont {Manceau}, \citenamefont
  {Bousseksou}, \citenamefont {Carusotto}, \citenamefont {Biasiol},
  \citenamefont {Colombelli},\ and\ \citenamefont
  {De~Liberato}}]{Cortese_2020}%
  \BibitemOpen
  \bibfield  {author} {\bibinfo {author} {\bibfnamefont {E.}~\bibnamefont
  {Cortese}}, \bibinfo {author} {\bibfnamefont {N.-L.}\ \bibnamefont {Tran}},
  \bibinfo {author} {\bibfnamefont {J.-M.}\ \bibnamefont {Manceau}}, \bibinfo
  {author} {\bibfnamefont {A.}~\bibnamefont {Bousseksou}}, \bibinfo {author}
  {\bibfnamefont {I.}~\bibnamefont {Carusotto}}, \bibinfo {author}
  {\bibfnamefont {G.}~\bibnamefont {Biasiol}}, \bibinfo {author} {\bibfnamefont
  {R.}~\bibnamefont {Colombelli}},\ and\ \bibinfo {author} {\bibfnamefont
  {S.}~\bibnamefont {De~Liberato}},\ }\bibfield  {title} {\bibinfo {title}
  {Excitons bound by photon exchange},\ }\bibfield  {journal} {\bibinfo
  {journal} {Nature Physics}\ }\href
  {https://doi.org/10.1038/s41567-020-0994-6} {10.1038/s41567-020-0994-6}
  (\bibinfo {year} {2020})\BibitemShut {NoStop}%
\bibitem [{\citenamefont {Seok}\ \emph {et~al.}(2011)\citenamefont {Seok},
  \citenamefont {Jamshidi}, \citenamefont {Kim}, \citenamefont {Dhuey},
  \citenamefont {Lakhani}, \citenamefont {Choo}, \citenamefont {Schuck},
  \citenamefont {Cabrini}, \citenamefont {Schwartzberg}, \citenamefont {Bokor},
  \citenamefont {Yablonovitch},\ and\ \citenamefont {Wu}}]{Seok-photonharvest}%
  \BibitemOpen
  \bibfield  {author} {\bibinfo {author} {\bibfnamefont {T.~J.}\ \bibnamefont
  {Seok}}, \bibinfo {author} {\bibfnamefont {A.}~\bibnamefont {Jamshidi}},
  \bibinfo {author} {\bibfnamefont {M.}~\bibnamefont {Kim}}, \bibinfo {author}
  {\bibfnamefont {S.}~\bibnamefont {Dhuey}}, \bibinfo {author} {\bibfnamefont
  {A.}~\bibnamefont {Lakhani}}, \bibinfo {author} {\bibfnamefont
  {H.}~\bibnamefont {Choo}}, \bibinfo {author} {\bibfnamefont {P.~J.}\
  \bibnamefont {Schuck}}, \bibinfo {author} {\bibfnamefont {S.}~\bibnamefont
  {Cabrini}}, \bibinfo {author} {\bibfnamefont {A.~M.}\ \bibnamefont
  {Schwartzberg}}, \bibinfo {author} {\bibfnamefont {J.}~\bibnamefont {Bokor}},
  \bibinfo {author} {\bibfnamefont {E.}~\bibnamefont {Yablonovitch}},\ and\
  \bibinfo {author} {\bibfnamefont {M.~C.}\ \bibnamefont {Wu}},\ }\bibfield
  {title} {\bibinfo {title} {Radiation engineering of optical antennas for
  maximum field enhancement},\ }\href {https://doi.org/10.1021/nl2010862}
  {\bibfield  {journal} {\bibinfo  {journal} {Nano Letters}\ }\textbf {\bibinfo
  {volume} {11}},\ \bibinfo {pages} {2606} (\bibinfo {year} {2011})},\ \bibinfo
  {note} {pMID: 21648393},\ \Eprint
  {https://arxiv.org/abs/https://doi.org/10.1021/nl2010862}
  {https://doi.org/10.1021/nl2010862} \BibitemShut {NoStop}%
\bibitem [{\citenamefont {Feuillet-Palma}\ \emph {et~al.}(2013)\citenamefont
  {Feuillet-Palma}, \citenamefont {Todorov}, \citenamefont {Vasanelli},\ and\
  \citenamefont {Sirtori}}]{Todorov-Thz-enhanc}%
  \BibitemOpen
  \bibfield  {author} {\bibinfo {author} {\bibfnamefont {C.}~\bibnamefont
  {Feuillet-Palma}}, \bibinfo {author} {\bibfnamefont {Y.}~\bibnamefont
  {Todorov}}, \bibinfo {author} {\bibfnamefont {A.}~\bibnamefont {Vasanelli}},\
  and\ \bibinfo {author} {\bibfnamefont {C.}~\bibnamefont {Sirtori}},\
  }\bibfield  {title} {\bibinfo {title} {Strong near field enhancement in {THz}
  nano-antenna arrays},\ }\href {https://doi.org/10.1038/srep01361} {\bibfield
  {journal} {\bibinfo  {journal} {Scientific reports}\ }\textbf {\bibinfo
  {volume} {3}},\ \bibinfo {pages} {1361} (\bibinfo {year} {2013})}\BibitemShut
  {NoStop}%
\bibitem [{\citenamefont {Seto}\ and\ \citenamefont
  {Helm}(1992)}]{helm-bistab}%
  \BibitemOpen
  \bibfield  {author} {\bibinfo {author} {\bibfnamefont {M.}~\bibnamefont
  {Seto}}\ and\ \bibinfo {author} {\bibfnamefont {M.}~\bibnamefont {Helm}},\
  }\bibfield  {title} {\bibinfo {title} {Charge transfer induced optical bistability 
in an asymmetric quantum well structure},\ }\href@noop {}
  {\bibfield  {journal} {\bibinfo  {journal} {Applied Physics Letters}\
  }\textbf {\bibinfo {volume} {60}},\ \bibinfo {pages} {859} (\bibinfo {year}
  {1992})}\BibitemShut {NoStop}%
\bibitem [{\citenamefont {Zaluzny}(1993)}]{Zaluzny-sat}%
  \BibitemOpen
  \bibfield  {author} {\bibinfo {author} {\bibfnamefont {M.}~\bibnamefont
  {Zaluzny}},\ }\bibfield  {title} {\bibinfo {title} {Saturation of intersubband 
absorption and optical rectification in asymmetric quantum wells},\ }\href {https://doi.org/10.1063/1.354339} {\bibfield  {journal}
  {\bibinfo  {journal} {Journal of Applied Physics}\ }\textbf {\bibinfo
  {volume} {74}},\ \bibinfo {pages} {4716} (\bibinfo {year}
  {1993})}\BibitemShut {NoStop}%
\bibitem [{\citenamefont {ZaÅ‚uÅ¼ny}\ and\ \citenamefont
  {Nalewajko}(1999)}]{zaluzny1999coupling}%
  \BibitemOpen
  \bibfield  {author} {\bibinfo {author} {\bibfnamefont {M.}~\bibnamefont
  {Zaluzny}}\ and\ \bibinfo {author} {\bibfnamefont {C.}~\bibnamefont
  {Nalewajko}},\ }\bibfield  {title} {\bibinfo {title} {Coupling of infrared
  radiation to intersubband transitions in multiple quantum wells: The
  effective-medium approach},\ }\href
  {https://doi.org/10.1103/PhysRevB.59.13043} {\bibfield  {journal} {\bibinfo
  {journal} {Phys. Rev. B}\ }\textbf {\bibinfo {volume} {59}},\ \bibinfo
  {pages} {13043} (\bibinfo {year} {1999})}\BibitemShut {NoStop}%
\bibitem [{\citenamefont {Drummond}(1981)}]{Drummond_bista}%
  \BibitemOpen
  \bibfield  {author} {\bibinfo {author} {\bibfnamefont {P.}~\bibnamefont
  {Drummond}},\ }\bibfield  {title} {\bibinfo {title} {Optical bistability in a
  radially varying mode},\ }\href {https://doi.org/10.1109/JQE.1981.1071100}
  {\bibfield  {journal} {\bibinfo  {journal} {IEEE Journal of Quantum
  Electronics}\ }\textbf {\bibinfo {volume} {17}},\ \bibinfo {pages} {301}
  (\bibinfo {year} {1981})}\BibitemShut {NoStop}%
\bibitem [{\citenamefont {Grant}\ and\ \citenamefont
  {Kimble}(1982)}]{Grant:82}%
  \BibitemOpen
  \bibfield  {author} {\bibinfo {author} {\bibfnamefont {D.~E.}\ \bibnamefont
  {Grant}}\ and\ \bibinfo {author} {\bibfnamefont {H.~J.}\ \bibnamefont
  {Kimble}},\ }\bibfield  {title} {\bibinfo {title} {Optical bistability for
  two-level atoms in a standing-wave cavity},\ }\href
  {https://doi.org/10.1364/OL.7.000353} {\bibfield  {journal} {\bibinfo
  {journal} {Opt. Lett.}\ }\textbf {\bibinfo {volume} {7}},\ \bibinfo {pages}
  {353} (\bibinfo {year} {1982})}\BibitemShut {NoStop}%
\bibitem [{\citenamefont {Mann}\ \emph {et~al.}(2021)\citenamefont {Mann},
  \citenamefont {Nookala}, \citenamefont {Johnson}, \citenamefont {Cotrufo},
  \citenamefont {Mekawy}, \citenamefont {Klem}, \citenamefont {Brener},
  \citenamefont {Raschke}, \citenamefont {Al\`{u}},\ and\ \citenamefont
  {Belkin}}]{Mann_Metasurface_2021}%
  \BibitemOpen
  \bibfield  {author} {\bibinfo {author} {\bibfnamefont {S.~A.}\ \bibnamefont
  {Mann}}, \bibinfo {author} {\bibfnamefont {N.}~\bibnamefont {Nookala}},
  \bibinfo {author} {\bibfnamefont {S.~C.}\ \bibnamefont {Johnson}}, \bibinfo
  {author} {\bibfnamefont {M.}~\bibnamefont {Cotrufo}}, \bibinfo {author}
  {\bibfnamefont {A.}~\bibnamefont {Mekawy}}, \bibinfo {author} {\bibfnamefont
  {J.~F.}\ \bibnamefont {Klem}}, \bibinfo {author} {\bibfnamefont
  {I.}~\bibnamefont {Brener}}, \bibinfo {author} {\bibfnamefont {M.~B.}\
  \bibnamefont {Raschke}}, \bibinfo {author} {\bibfnamefont {A.}~\bibnamefont
  {Al\`{u}}},\ and\ \bibinfo {author} {\bibfnamefont {M.~A.}\ \bibnamefont
  {Belkin}},\ }\bibfield  {title} {\bibinfo {title} {Ultrafast optical
  switching and power limiting in intersubband polaritonic metasurfaces},\
  }\href {https://doi.org/10.1364/OPTICA.415581} {\bibfield  {journal}
  {\bibinfo  {journal} {Optica}\ }\textbf {\bibinfo {volume} {8}},\ \bibinfo
  {pages} {606} (\bibinfo {year} {2021})}\BibitemShut {NoStop}%
\end{thebibliography}
\end{document}